\documentstyle[twocolumn,psfig]{article}
\begin{document}
\renewcommand{\textfraction}{0}
\renewcommand{\topfraction}{1}
\renewcommand{\bottomfraction}{1}
\rule[-8mm]{0mm}{8mm}
\twocolumn[{\begin{center}
{\large\bf Dynamical Magnetic Susceptibility for the $t$-$J$ Model}\\[4mm]
Th.~Pruschke, Th.\ Obermeier and J.\ Keller\\[4mm]
Institut f\"ur Festk\"orperphysik, Universit\"at Regensburg, 93040
Regensburg, Germany
\end{center}}

\vspace*{0.5cm}
We present results for the {\em dynamical}\/ magnetic susceptibility of the 
$t$-$J$ model, calculated with the dynamical mean field theory. For $J=0$
we find enhanced ferromagnetic correlations but an otherwise relatively
$\vec{q}$-independent dynamical magnetic susceptibility. For $J>0$ 
the explicit antiferromagnetic exchange leads to a dynamic spin structure
factor with the expected peak at the antiferromagnetic Bragg point.\vskip0.75cm]
\normalsize

\vspace{-0.175cm}
The observation that the local electron-electron interaction in models 
of strongly correlated electronsystems leads in the limit
spatial dimensions $d\to\infty$ to a local one-particle self energy \cite{metzvoll},
founded the  so-called {\em dynamical mean field theory} (DMFT): The 
lattice problem maps onto an impurity system coupled to an
effective, self-consistently determined bath \cite{locapprox},
the lattice structure enters only via the free density of states (DOS)
$\rho_0(\epsilon)=\sum\limits_{\vec{k}}\delta(\epsilon-\epsilon_{\vec{k}})$.

In addition to the calculation of one-particle properties the DMFT also allows
to obtain the two-particle quantities
\cite{jarrell}, especially the {\em dynamical}\/ magnetic susceptibility.
So far, however, only static susceptibilities have been studied \cite{jarrell,jafre,pruqi}.
In this contribution we want to present first results for the dynamical
magnetic susceptibility of a strongly correlated electron system calculated within the DMFT.

The model we want to discuss is the well-known $t$-$J$ model
$$
H_{\rm tJ} = -t\sum\limits_{\langle i,j\rangle\sigma} 
\hat{c}_{j\sigma}^\dagger \hat{c}_{i\sigma}
+J\sum\limits_{\langle i,j\rangle} \vec{S}_i\cdot\vec{S}_j\;\;.
$$
The operator $\hat{c}_{i\sigma}^{\dagger}$ creates an electron with spin
$\sigma$ on site $i$ if and only if no other particle is present, the $\vec{S}_i$
are the conventional spin operators and $\langle\cdot,\cdot\rangle$ denotes sums
on nearest neighbours only.
In a previous publication \cite{pruqi}, we have shown that the magnetic
susceptibility for this model acquires a particularly simple form, namely
$\chi^{tJ}_{\vec{q}}(z)=\chi^\infty_{\vec{q}}(z)\left[1-J_{\vec{q}}\chi^\infty_{\vec{q}}(z)\right]^{-1}$. Here, $\chi^\infty_{\vec{q}}(z)$ denotes the magnetic
susceptibility for $J=0$, i.e.\ the $U=\infty$ Hubbard model, and $J_{\vec{q}}$
is given by
\newpage
$$
J_{\vec{q}}=-\frac{J}{d}\sum\limits_{l=1}^{d}\cos(q_l\cdot a)\;\;.
$$

The major problem is the calculation of $\chi^\infty_{\vec{q}}(z)$. 
This task
can be performed most conveniently in Matsubara space, since there the calculation
reduces to the inversion of matrices. The final step, the analytic continuation
$\chi^\infty_{\vec{q}}(i\nu_n)\to\chi^\infty_{\vec{q}}(\omega+i\delta)$ is then
performed with Pad\'e approximation. To calculate $\chi^\infty_{\vec{q}}(z)$ we
use standard resolvent techniques in connection with the NCA \cite{nca} as explained in reference \cite{pruqi}.
\begin{figure}[hb]
\centerline{\psfig{figure=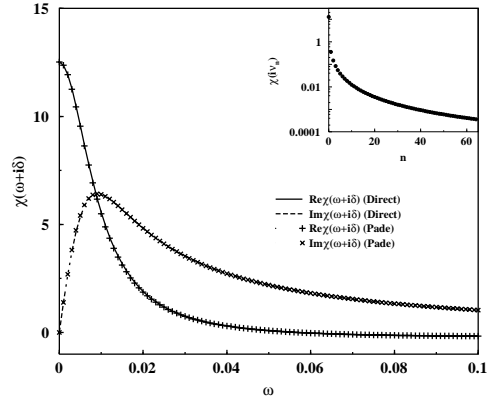,width=0.4\textwidth}\ }
\caption[]{Local dynamical magnetic susceptibility for $J=0$
obtained directly from the effective local model (lines) and
from the Matsubara data via numerical analytic continuation with Pad\'e approximants
(symbols). The inset shows the underlying data in Matsubara space.
The parameters in the calculation are $\delta=5$\% and $T=t/30$.}
\end{figure}
Note that since we solve the DMFT by perturbative techniques the results for $\chi^\infty_{\vec{q}}(i\nu_n)$ do not contain statistical errors like in QMC
and thus the analytic continuation via Pad\'e approximants is expected to work properly.

Let us nevertheless start with some ``confidence tests'' for the latter procedure.
This can most conveniently be done with the help of the local susceptibility,
where one can obtain both, $\chi_{loc}(\omega+i\delta)$ and
$\chi_{loc}(i\nu_l)$ {\em independently}.
In Fig.~1 we compare directly calculated results for the local dynamic
susceptibility (full lines) to those obtained from a Pad\'e approximation
(symbols).
The typically rather structureless data on the imaginary axis are displayed in 
the inset.
Note that we used merely $n=64$ Matsubara frequencies in the Pad\'e approximation.
This example shows that at least for mildly varying functions like susceptibilities
the Pad\'e approximation is a rather accurate tool for numerical analytic continuation.

\begin{figure}[htb]
\centerline{\psfig{figure=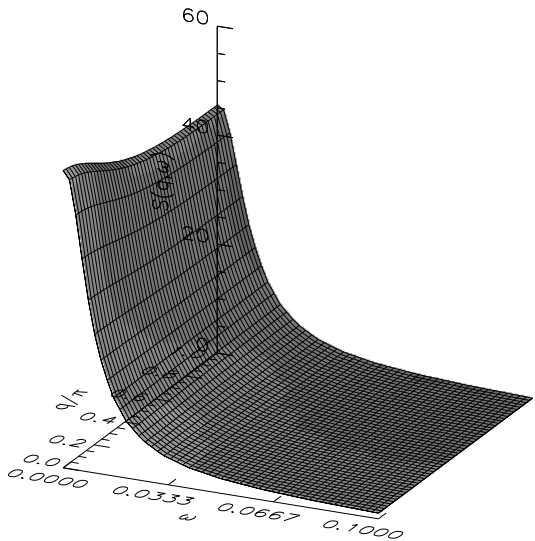,width=0.6\textwidth}\ }
\caption[]{Spin strucure factor at $J=0$ for
$\delta=5$\% and $T=t/30$ as function of $\vec{q}$ an $\omega$. Most
prominent is a slight enhancement of {\em ferromagnetic} fluctuations for
$\omega\to0$ but an otherwise rather weak variation with $\vec{q}$.}
\end{figure}
The complete dynamical spin-structure factor ($g(\omega)$ denotes the Bose function)
$S(\vec{q},\omega)=
\Im m\left\{\chi^\infty_{\vec{q}}(\omega+i0^+)\right\}\cdot\left\{1+g(\omega)\right\}$
for $J=0$, i.e.\ the $U=\infty$ Hubbard model, 
is shown in Fig.~2. The doping is $\delta=5$\% and the temperature $T=t/30$.
The result is quite characteristic for the spin correlations in this model:
As in the static case \cite{pruqi}, we find a rather weak $\vec{q}$-dependence
of $S(\vec{q},\omega)$. An interesting aspect is that for $\omega\to0$ the
ferromagnetic correlations at $q=0$ are enhanced, while for increasing energy
$\vec{q}=(\pi,\pi,\ldots)$ slightly dominates. This result can be understood
by the fact that, since there is no net magnetic exchange present in this
model, no special configuration is preferred by the system, except for
$\omega\to0$ where a small gain in kinetic energy by a partially ferromagnetic
polarization \cite{nagaoka} is responsible for enhanced ferromagnetic correlations. 

\begin{figure}[htb]
\centerline{\psfig{figure=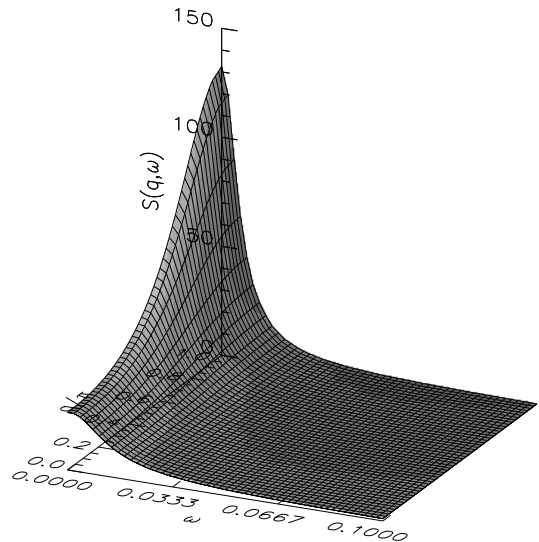,width=0.6\textwidth}\ }
\caption[]{Spin structure factor for the $t$-$J$ model with $J=T$, other
parameters as in Fig.~2. The fluctuations are now strongest for $\omega\to0$
and $\vec{q}$ at the antiferromagnetic Bragg point, as expected.}
\end{figure}
The situation of course changes when we include a finite exchange $J=T$: As is
clear from Fig.~3, the antiferromagnetic correlations are strongly enhanced,
as expected.
It is noteworthy that the weak $\vec{q}$-dependence in
$\chi^\infty_{\vec{q}}(\omega+i0^+)$ 
has no significant influence for larger $J$. Note that the value
of $J$ chosen here is still only $50$\% of the critical value where a transition
into an ordered state occurs for the current parameter values. This supports
our earlier suggestion that the prominent effects of the spin dynamics for the
$t$-$J$ model in the paramagnetic phase can already be addressed by replacing
$\chi^\infty_{\vec{q}}(\omega+i0^+)\to\chi_{loc}(\omega+i0^+)$
\cite{pruqi}.

To conclude we presented DMFT-results for the dynamical susceptibility for
the $t$-$J$ model. We showed that for
the present problem the numerical analytic continuation by means of Pad\'e approximation
is a reliable tool. The results for the dynamical susceptibility of the
$U=\infty$ Hubbard model show a behaviour as already conjectured from the
statics \cite{pruqi}, namely a rather weak variation with $\vec{q}$ and a
slight enhancement of the ferromagnetic correlations as $\omega\to0$. A finite 
exchange interaction $J$ again strongly enhances the antiferromagnetic
fluctuations. In this case the detailed $\vec{q}$ structure of the underlying
susceptibility of the Hubbard model is completely washed out. This observation
gives to some extent a microscopic justification of recent suggestions for
approximate forms of the dynamical magnetic susceptibility for strongly
correlated models \cite{pines}.

The authors would like to thank Dr.\ A.\ Liechtenstein for providing the code
for the Pad\'e approximations.
This work was supported by the Deutsche Forschungsgemeinschaft
grant number \mbox{Pr 298/3-1}.

\end{document}